# Alternative Model of Space-Charge Limited Thermionic Current Flow Through a Plasma

M. D. Campanell (michaelcampanell@gmail.com)
*Lawrence Livermore National Laboratory, Livermore, CA 94551, USA*

It is widely assumed that thermionic current flow through a plasma is limited by a "space-charge limited" (SCL) cathode sheath that consumes the hot cathode's negative bias and accelerates upstream ions into the cathode. Here, we formulate a fundamentally different current-limited mode. In the "inverse" mode, the potentials of both electrodes are above the plasma potential, so that the plasma ions are confined. The bias is consumed by the anode sheath. There is no potential gradient in the neutral plasma region from resistivity or presheath. The inverse cathode sheath pulls some thermoelectrons back to the cathode, thereby limiting the circuit current. Thermoelectrons entering the zero-field plasma region that undergo collisions may also be sent back to the cathode, further attenuating the circuit current. In planar geometry, the plasma density is shown to vary linearly across the electrode gap. A continuum kinetic planar plasma diode simulation model is set up to compare the properties of current modes with classical, conventional SCL, and inverse cathode sheaths. SCL modes can exist only if charge-exchange collisions are turned off in the potential well of the virtual cathode to prevent ion trapping. With the collisions, the current-limited equilibrium must be inverse. Inverse operating modes should therefore be present or possible in many plasma devices that rely on hot cathodes. Evidence from past experiments is discussed. The inverse mode may offer opportunities to minimize sputtering and power consumption that were not previously explored due to the common assumption of SCL sheaths.

## I. Introduction

Limitation of thermionic current by the space charge effect is an important phenomenon that has been studied for a century [1]. There are still open research questions [2,3], but the basic cause of current limitation is well understood in vacuum. When the thermoelectrons reach a high enough density, their negative space charge repels some back to the cathode. Current limitation in a planar vacuum diode is reviewed by simulations in Fig. 1. Panel (a) shows that all thermoelectrons contribute to the current until the Child-Langmuir (CL) [4] emission threshold is reached, beyond which the current saturates. The profiles of potential $\varphi(x)$ in Fig. 1(b) show the formation of the "virtual cathode" (VC) barrier responsible for the saturation. Electron distribution functions for emission intensities below and beyond the CL threshold are compared in Fig. 1(c,d).

Thermionic current flow can be enhanced by introducing plasma between the electrodes. The positive ions offset the negative space charge, allowing more electrons into the gap before saturation occurs. Plasma diodes are more complex and less understood than vacuum diodes. One must consider how the plasma electrons, ions, and thermoelectrons interact with each other, the self-consistent electric field, neutrals, and the electrodes. This is crucial for understanding the current flow, power dissipation and electrode erosion for any plasma application that relies on hot cathodes. Examples include thermionic discharges[5,6], thermionic converters[7], thermionic tethers[8], emissive probes[9], arc cutting [10], torches [11] and electron transpiration cooling [12].

An early treatment of the hot cathode sheath with ions was in Langmuir's 1929 paper [13]. Langmuir extended the vacuum diode model by adding a current of ions flowing opposite to the thermoelectrons. He argued that when the thermionic emission reaches a critical intensity a VC begins to form, as in vacuum. Many other authors including Prewett and Allen (1976) [14], Takamura *et al.* (2004) [15], Gyergyek and Čerček (2007) [16], Din (2013) [17], Pekker and Hussary (2015) [18], and Cavalier *et al.* (2017) [9] extended Langmuir's model. Their solutions known as "space-charge limited" (SCL) sheaths are used when predicting how plasmas interact with hot cathodes in current-limited conditions. SCL type sheaths were also predicted to form at floating surfaces with strong electron emission [19,20,21,22,23,24].

A serious problem with the conventional SCL sheath theory was recently revealed. In practice, some of the positive ions passing through the potential well of the VC will suffer charge-exchange (CX) collisions with cold neutral atoms, leaving behind cold ions which become trapped in the well. Campanell and Umansky showed that the accumulating ions in floating [25] and biased [26] SCL sheaths will force the sheath to transition to the "inverse" shape, such that the plasma potential is below the surface potential. In an inverse equilibrium, ions are confined and the upstream plasma structure differs from the assumptions behind SCL theories.

The physics of current flow in the hot cathode inverse mode is not yet explained. But experimental evidence of the mode has long existed. The "anode glow mode" (AGM) of thermionic discharges [27,28] has properties suggesting an inverse cathode sheath. In the 1950's, Malter *et al.* [27] noted that since visible light appeared only near the anode, the cathode sheath cannot be the SCL type that Langmuir proposed, or else thermoelectrons would accelerate to excitation energies closer to the cathode. Yet in today's research it is still assumed that emitting sheaths are SCL when some of the emission is suppressed [29]. Since CX ion trapping is now known to inhibit SCL sheaths [25,26], *every* hot cathode-plasma system operating in current limited conditions is probably in an inverse mode. This may lead to unexpected behaviors and unexplored operating regimes of many devices.

A formal theory is needed to describe how the plasma and electrode sheaths couple to each other in the inverse mode. In Section II, we discuss the fundamental properties of the classical, SCL and inverse current modes. Section III provides an analytical model of the inverse mode in planar geometry. Section IV presents continuum kinetic simulations of a plasma between a hot cathode and anode. Modes with classical, SCL and inverse cathode sheaths are demonstrated and compared. Section V studies plasma instabilities that arise in the inverse mode. Further discussion of the inverse mode is in Section VI, followed by conclusions in Section VII.

## II. Fundamental Aspects of the Hot Cathode Current Modes

Understanding thermionic current flow requires answering a fundamental question, "*How does the potential get from A to B in a plasma?*"[30]. In general, there are sharp potential variations in the thin sheaths near surfaces, and more gradual variations across the quasineutral region. Here a "sheath potential difference" is defined as the surface potential relative to its sheath edge's potential. Thus it is negative for classical Debye sheaths. Many models of hot cathode sheaths have been constructed [14,15,16,17] where the cathode sheath potential difference is an input parameter assumed negative. But in practice only the potential of the cathode relative to the anode is constrained to be negative by a bias. It turns out the cathode sheath potential difference can take either sign depending on how the potential distributes itself across the gap. In this section we will outline the possible distributions, which are sketched in Fig. 2.

### A. Cathode Sheath Physics

One may expect a negatively biased cathode to acquire a negative (ion-accelerating) sheath potential difference. When the sheath is classical as sketched in Fig. 2, the total charge in the sheath is positive. The Bohm criterion [31] stipulates that the ions need to enter the sheath with a substantial flow velocity to set up the positive space charge layer. A "presheath" electric field in the plasma is then needed to accelerate the ions to the Bohm velocity at the sheath edge. Thermionic emission may create a layer of net negative charge near the cathode. As long as that layer is weaker than the positive charge layer, the sheath remains classical monotonic, like the sheath at a cold cathode [32,33]. The entire thermionic flux, determined by the hot cathode's temperature, is able to enter the plasma. Hence the classical sheath mode is also called a "temperature-limited mode" (TLM) [44].

Langmuir [13] first argued that if the emission intensity surpasses the maximum value compatible with a classical monotonic sheath, *"the current becomes limited by space charge and a further increase in electron current cannot occur. The cathode is then covered by a double layer or double sheath..."* Langmuir demonstrated the current saturation in a diode model where thermoelectrons and ions passed through the gap in opposite directions. Later researchers modeling hot cathode sheaths [15,17,34] improved upon Langmuir's model and accounted for the Bohm criterion. They showed that as the emitted flux $\Gamma_{emit}$ is raised in the classical sheath mode, the electric field at the cathode surface weakens and eventually vanishes. Further increases of $\Gamma_{emit}$ are assumed to be suppressed by formation of a small VC, as in the "space-charge limited" (SCL) sheath sketched in Fig. 2.

In a SCL mode, the overall cathode sheath potential difference is still negative. Hence Langmuir predicted [13], *"... use of a hot cathode, no matter how high its temperature, does not destroy the cathode drop."* But recent work shows that the inevitable accumulation of charge-exchange ions in the VC potential well may in fact destroy the cathode drop, forming an inverse mode [26]. In the inverse mode sketched here in Fig. 2, the cathode sheath is a single layer of negative charge formed by thermoelectrons. Ions are confined by the potential distribution and their density drops towards zero in both sheaths.

**B. Electric Fields in the Plasma**

Since $\varphi(x)$ decreases from the cathode to the cathode sheath edge in the inverse mode, it must rise elsewhere to reach the bias on the anode. An important question is, how much of the bias is consumed by the plasma, and how much by the anode sheath?

When Malter, Johnson and Webster [27] studied the anode glow mode of thermionic discharges, they sketched a qualitative prediction of the potential distribution. It is similar to our sketch of the inverse mode in Fig. 2, except theirs shows $\varphi(x)$ rising across the interior. They stated, *"The slow rise through the plasma is required to impart sufficient directed velocity to the electrons in the plasma to sustain the circuit current."* In other words, they expected a potential rise from "resistivity". However, that would require there to be a local potential minimum near the cathode, and the cold trapped ions would all settle to that one point, which is impossible [26]. For that reason, we claim $\varphi(x)$ must be flat between the sheaths in an inverse mode. The potential can only rise in the anode sheath (where there are no trapped ions).

To prove that a resistivity field is unnecessary in the inverse mode, it is worthwhile to discuss why it is necessary in the other modes. When the cathode sheath is classical or SCL, the injected thermoelectrons can only escape at the anode. Collisions in the plasma region may redirect some thermoelectrons back towards the cathode. But since most collisions will deplete some electron energy or scatter it in transverse dimensions, the probability of overcoming the cathode sheath barrier is low. For global charge balance, the thermoelectrons must escape at the anode at the same net rate they enter the plasma (neglecting sinks or losses to other surfaces). Since collisions act to isotropize electron velocities, an electric field must form to maintain a spatially-conserved current directed towards the anode.

Overall, the resistivity field forms in the classical and SCL modes because the electrons are unable to pass from the plasma back to the cathode. That is an implicit assumption in many theoretical models of plasma diodes. For example, the model by Pekker and Hussary [18] treats the hot cathode sheath, plasma, and anode sheath, patched consistently. They assume the full thermionic flux contributes to the circuit current when the sheath is classical (or if the sheath is SCL, the SCL flux sets the current). That current, along with the plasma conductivity, determines the required resistivity field.

We can now answer why a resistivity field is not needed to drive electrons towards the anode in the inverse mode. It is because there is no potential barrier blocking them from escaping to the cathode. In Sec. III.B we will show that most thermoelectrons entering the neutral region that undergo collisions will in fact return to the cathode. Thus the current in the inverse mode is limited not only by the "space charge" potential barrier of the cathode sheath but also by collisional redirection within the plasma.

**C. Anode Sheath Physics**

For a plasma between biased electrodes, the usual expectation is that the anode sheath potential difference is negative [35]. In that case the full applied bias has to be consumed by the cathode sheath and plasma resistivity. Normally the anode sheath potential difference cannot be positive since the high loss rate of electrons would exceed the global loss rate of ions. Exceptions to this rule are when the anode is too small to affect global current balance [35], the plasma source is electron-rich [32], or the cathode emits electrons at a rate sufficient to compensate rapid losses at the anode. Indeed there is research on electric arcs suggesting that the anode sheath potential difference can be either sign [36,37]. We will confirm in simulations that it can take either sign when the hot cathode sheath is classical or SCL. Meanwhile in the inverse mode, it is always positive.

**III. THE INVERSE MODE – ANALYTICAL MODEL**

Let us consider a planar inverse mode with electrode gap size L and mutual bias $V_{bias}$. The positive ions are confined and assumed to have zero temperature. A thermionic flux $\Gamma_{emit}$ with temperature $T_{emit}$ is emitted from the cathode with a half-Maxwellian velocity distribution $\propto \exp(-m_e v_x^2/2T_{emit})$ for $v_x > 0$. Emission from the anode is not considered since it would get reflected promptly back to the anode in the inverse mode. (In classical and SCL modes, emission from the anode can enter the plasma and play a significant role discussed by Levko [38] and Khrabry *et al*. [39].)

**A. The collisionless inverse mode**

We first treat the situation where the electron mean free path is long enough to neglect collisions. The structure of the collisionless inverse mode is sketched in Fig. 3. The inverse sheath potential difference (cathode potential relative to the inverse sheath edge) is positive, denoted $\Phi_{inv}$, see Fig. 3(a). The unit charge $q_e$ is considered a positive quantity here. The inverse sheath is a potential barrier to emitted thermoelectrons. It is known [40] that half-Maxwellian particles entering a potential barrier remain half-Maxwellian while decreasing in density such that the injected (+x-directed) flux reaching the inverse sheath edge must be,

$$\Gamma_{emit} \exp\left(\frac{-q_e \Phi_{inv}}{T_{emit}}\right) = \Gamma_e \qquad (1)$$

In the collisionless case the plasma-injected flux above transits unimpeded towards the anode sheath and thus is the net flux $\Gamma_e$. The neutral plasma region is half-Maxwellian electrons propagating through the background of cold trapped ions with a density $N_p$ that must be uniform, see Fig. 3(c). By integrating the half-Maxwellian distribution times $v_x$ [40], the net flux $\Gamma_e$ in terms of density and temperature is found,

$$\Gamma_e = N_p \sqrt{\frac{2T_{emit}}{\pi m_e}} \qquad (2)$$

The hot cathode temperature $T_{emit}$ is often controlled or measurable in experiments. In terms of $T_{emit}$ and the material's work function, $\Gamma_{emit}$ is calculable according to the Richardson-Dushman equation [1]. One may also determine $\Gamma_{emit}$ as a function of $T_{emit}$ experimentally by measuring current in a temperature-limited mode. Then in a collisionless inverse mode, the $\Gamma_e$ determined by measured current can be used to calculate $\Phi_{inv}$ and $N_p$ with Eqs. (1) and (2).

The plasma in the collisionless inverse mode can carry the maximum thermionic current density $q_e\Gamma_{emit}$ if $\Phi_{inv}$ is small. In that limiting case the plasma density equals the density of emitted electrons at the cathode surface, a known quantity $N_{emit} = \Gamma_{emit}(\pi m_e/2T_{emit})^{1/2}$. A useful relationship between $\Phi_{inv}$ and the density ratio $N_{emit}/N_p$ is found after combining (1) with (2),

$$\Phi_{inv} = \frac{T_{emit}}{q_e} \ln\left(\frac{N_{emit}}{N_p}\right) \qquad (3)$$

Let us now discuss the sheath structures. The ion density is zero in both sheaths. The electric field is zero at both sheath edges. The anode sheath is like a vacuum diode at the CL saturation threshold. Electrons entering it get accelerated across a potential drop $V_{bias} + \Phi_{inv}$. Because $q_e\Phi_{inv} \sim T_{emit}$ via (3) and assuming $T_{emit} \ll q_eV_{bias}$, initial velocities at the anode sheath edge are unimportant and the total drop is comparable to $V_{bias}$. So one can invoke the familiar expression for the maximum planar CL current density [1] to show that $q_e\Gamma_{CL} = (2^{5/2}\varepsilon_0/9)(q_e/m_e)^{1/2}(V_{bias}^{3/2}/L_{ash}^2)$, where $L_{ash}$ is the anode sheath size. Then matching the flux to our Eq. (2) leads to an expression for $L_{ash}$,

$$L_{ash} = \frac{2\varepsilon_0^{1/2}\pi^{1/4}}{3q_e^{1/4}} \cdot \frac{V_{bias}^{3/4}}{N_p^{1/2}T_{emit}^{1/4}} \qquad (4)$$

The inverse cathode sheath contains thermoelectrons moving away from the cathode and ones that reflect within the sheath. No electrons enter the cathode sheath from the plasma in the collisionless case. The electron density is expressible in terms of the potential $\varphi$ relative to the cathode, $n_e(\varphi) = N_{emit}\exp(q_e\varphi/T_{emit})\{1+\text{erf}[(q_e(\Phi_{inv}+\varphi)/T_{emit})^{1/2}]\}$. But the spatial profiles of $n_e(x)$ and $\varphi(x)$ can only be obtained by numerical solution of Poisson's equation.

Since $n_e(x) > N_p$ at all x within the cathode sheath, a useful upper limit on its size $L_{csh}$ can be derived by supposing $n_e$ were flat with density $N_p$. Solving for the parabolic $\varphi(x)$ with the sheath potential difference $\Phi_{inv}$ and zero electric field at the sheath edge used as boundary conditions leads to,

$$L_{csh} < \sqrt{\left(\frac{2\varepsilon_0 T_{emit}}{q_e^2}\right)\left(\frac{\ln(N_{emit}/N_p)}{N_p}\right)} \qquad (5)$$

Generally, the cathode sheath will be thin relative to the gap and $\Phi_{inv}$ limited to a few $T_{emit}$. Although (3) and (5) suggest the cathode sheath may become large when $N_p \ll N_{emit}$, the logarithms limit the sheath growth and the inverse mode breaks down at very low $N_p$ for a reason explained below.

An interesting comparison is that φ(x) in the collisionless inverse mode is like a current-limited Child-Langmuir solution (Fig. 1(b)) with an extended potential minimum. Extending the minimum is equivalent to reducing the electrode gap (which allows more current to flow). From this we conclude that $\Gamma_e$ in a collisionless inverse mode always exceeds $\Gamma_{CL}$. Also, since the extended potential minimum is the neutral region, $N_p$ must be at least as large as the electron density at the potential minimum of the CL solution, which is $\Gamma_{CL}(\pi m_e/2T_{emit})^{1/2}$, in order for an inverse mode to exist. Furthermore, we can conclude that $L_{csh}$ is always less than the distance from the cathode to the potential minimum of the CL solution.

**B. The collisional inverse mode**

Let us now consider how electron collisions affect the inverse mode. One important consequence is that some thermoelectrons injected into the plasma will change direction and fall back through the cathode sheath, contributing no net current to the circuit. Fewer electrons reach the anode sheath edge, making the density there lower than at the cathode sheath edge, as sketched in Fig. 4.

The dominant electron scattering process in most hot cathode devices is collisions with neutral atoms. In an inverse mode, the thermoelectrons entering the neutral region have energies $\sim T_{emit} < 1$eV, insufficient to excite or ionize any gas. So we assume the collisions are elastic with a known mean free path $e_{mfp}$ calculable from the background neutral density and cross sections. Because energy exchange in elastic collisions between light and heavy particles is known to be inefficient [41], we neglect it and assume that thermoelectrons keep their initial energy.

Given that the electric field in the plasma region is zero in the inverse mode, the motion of the electrons is altered just by collisions. They essentially undergo a random walk until they enter one of the absorbing sheaths. The thermoelectrons injected from the inverse sheath edge are assumed to be half-Maxwellian over $v_x$ and full Maxwellian in the transverse directions $\{v_y, v_z\}$ with temperature $T_{emit}$. Assuming collisions are isotropic, random, and energy-conserving, an ensemble of injected electrons becomes full Maxwellian with the same $T_{emit}$ after a collision. Accounting for injection and losses through sheaths, the overall electron velocity distribution can never be full Maxwellian in the plasma. The distribution over $v_x$ at each x takes the form of two half-Maxwellians with unequal densities $n_l(x)$ and $n_r(x)$, as sketched in Fig. 4.

Let us define x=0 and x=$L_p$ as the plasma boundaries. We wish to calculate as a function of x the x-directed flux of electrons moving left $\Gamma_l$ and right $\Gamma_r$ (both considered positive, such that $\Gamma_e = \Gamma_r - \Gamma_l$). If electrons moved only in the *x* dimension, the probability that an electron with a given positive $v_x$ passing through a thin region dx suffers an isotropic collision that changes its x-direction is ½dx/$e_{mfp}$. When transverse velocity is accounted for, the probability is doubled based on the fact that the average of $(v_x^2+v_y^2+v_z^2)^{1/2}$ over a Maxwellian distribution is twice the average of $|v_x|$ [40], thus doubling the actual average speed traveled through the neutral gas.

In terms of fluxes, we conclude at a given point, $\Gamma_r(x+dx)$ will equal $\Gamma_r(x)$ minus the portion of these electrons that redirect in the region dx, plus a contribution from leftward electrons at (x+dx) that redirect in the same region. That is,

$$\Gamma_r(x+dx) = \Gamma_r(x) - \Gamma_r(x) \cdot \frac{dx}{e_{mfp}} + \Gamma_l(x+dx) \cdot \frac{dx}{e_{mfp}} \quad (6)$$

Eq. (6) in the limit dx → 0 leads to a differential equation for $\Gamma_r$. A similar derivation for $\Gamma_l$ turns out to have an identical right hand side.

$$\frac{d\Gamma_r}{dx} = \frac{-\Gamma_r + \Gamma_l}{e_{mfp}}$$
$$\frac{d\Gamma_l}{dx} = \frac{\Gamma_l - \Gamma_r}{e_{mfp}} \quad (7)$$

From (7) we obtain two equations with integration constants $c_a$ and $c_b$ to be determined by boundary conditions.

$$\Gamma_l(x) = \Gamma_r(x) + c_a$$
$$\Gamma_r(x) = \frac{c_a}{e_{mfp}} x + c_b. \quad (8)$$

At the cathode sheath edge (x=0), $\Gamma_r = \Gamma_{emit}\exp(-q_e\Phi_{inv}/T_{emit}) = c_b$. At the anode sheath edge (x=$L_p$), $\Gamma_l$ must be zero, assuming all electrons entering the anode sheath fall into the anode. That is valid when collisions within the anode sheath are negligible. With collisions, that is still a good approximation provided the potential drop one $e_{mfp}$ into the sheath is enough that a typical collision redirects too much x-energy to the transverse dimensions, prohibiting a return to the anode sheath edge.

Solving for $c_a$ gives -$\Gamma_{emit}\exp(-q_e\Phi_{inv}/T_{emit})/(1+L_p/e_{mfp})$. Now $\Gamma_r$ and $\Gamma_l$ are known. Their sum is proportional to the total number of electrons $n_l + n_r$ at each x (this relies on the assumption of energy conservation in collisions resulting in uniform electron temperature $T_{emit}$ across the plasma). From that we determine the spatial distribution of plasma density $N_p(x)$ to be linear in x.

$$N_p(x) = N_{emit} \exp\left(\frac{-q_e\Phi_{inv}}{T_{emit}}\right) \cdot \left(\frac{2L_p + e_{mfp} - 2x}{L_p + e_{mfp}}\right) \quad (9)$$

It may be surprising that the ions were not considered in deriving $N_p(x)$. In the inverse mode, the ions are just a confined background of positive charge whose purpose is to neutralize whatever profile the electrons take. The electrons obey (9) when the electric field is negligible, which holds in the cold ion limit. If $T_i > 0$, there will be a pressure gradient $d(n_iT_i)/dx$ and a nonzero electric field will have to exist to offset it. The self-consistent effect of the field and density distribution on each other is beyond the scope of this analytical model but will be illustrated later by our simulations, see Fig. 11.

The net flux $\Gamma_e$ follows by putting x = $L_p$ into (9), since all electrons at the anode sheath edge are flowing into the anode,

$$\Gamma_e = \Gamma_{emit} \exp\left(\frac{-q_e\Phi_{inv}}{T_{emit}}\right) \cdot \left(\frac{e_{mfp}}{L_p + e_{mfp}}\right) \quad (10)$$

The exponential factor in (10) is the attenuation of current by the inverse sheath. The $e_{mfp}/(L_p+e_{mfp})$ factor is the current attenuation attributable to collisions. When $e_{mfp} \ll L_p$, most thermoelectrons return to the cathode even if the inverse sheath is weak. The current is not guaranteed to exceed the CL current that flows through a vacuum diode with the same {$V_{bias}$, $T_{emit}$, $\Gamma_{emit}$, L}.

Equations (9) and (10) allow one to determine $N_p(x)$, $\Gamma_e$ and $\Phi_{inv}$ in terms of parameters that are controllable in experiments {$\Gamma_{emit}$, $T_{emit}$, $e_{mfp}$, $L_p$}. (L is the controlled quantity but assuming the plasma fills most of the gap it is reasonable to approximate $L_p \approx L$.) Since there are three unknowns and two equations, an additional piece of information is needed to determine the system state. Measuring either the circuit current or the plasma density at one point is sufficient.

Using (9) with x=0 and x=$L_p$, the ratio of plasma density at the cathode sheath edge $N_{cse}$ to the anode sheath edge $N_{ase}$ is found.

$$\frac{N_{cse}}{N_{ase}} = \frac{2L_p + e_{mfp}}{e_{mfp}} \quad (11)$$

The sheaths in the collisional inverse mode are similar in structure to the collisionless case. If we neglect collisions within the anode sheath, it is still like a vacuum diode, so Eq. (4) with $N_p$ replaced by $N_{ase}$ estimates $L_{ash}$. That estimate is only an upper limit since collisions make electrons spend a longer time in the anode sheath, leading to a higher $n_e$ and a thinner sheath. Putting $N_{cse}$ in place of $N_p$ in (5) gives an upper bound on $L_{csh}$.

The maximum possible plasma density in the inverse mode is obtained by putting $\Phi_{inv} \to 0$ in (9). At high collisionality ($e_{mfp} \ll L_p$) the maximum $N_{cse}$ is $2N_{emit}$, twice the maximum density allowed in the collisionless case.

## IV. Simulation Study:

We developed a new simulation model to study thermionic current flow through plasma. The code design was motivated by a few issues. Many past computational studies of hot cathode sheaths simulated a collisionless domain between a cathode and a source boundary where the plasma ions and electrons were injected [15,16]. Such boundary injection models suffer from source sheath problems [42,43] and cannot capture the key influences of the presheaths, resistivity field and anode sheath.

To study current flow realistically, the entire plasma between a cathode and anode must be simulated, with ion and electron collisions included. Some 1D particle-in-cell (PIC) codes by previous authors simulated a planar thermionic discharge with rigorous collision operators [44,45]. Discharge modes consistent with a classical sheath mode (TLM) and inverse mode (AGM) were observed. The causes of mode transitions between the TLM and AGM were demonstrated for the first time by our recent thermionic discharge simulations in Ref. [26].

Thermionic discharge simulations are not ideal for the purposes of this paper due to their complex parameter dependencies. For example, changing a discharge parameter such as neutral density changes the plasma density and the rates of e-n and i-n collisions. Also, the plasma density in a thermionic discharge rises with $\Gamma_{emit}$ such that a current-limited mode may not be reachable by simply raising $\Gamma_{emit}$.

So here we will create a simpler plasma diode simulation model where the emitted flux, plasma density, temperature, and collision rates are controlled independently. This will allow us to focus on the basic properties of the plasma and sheaths in each current mode in terms of fundamental parameters.

### A. Continuum Kinetic Code

Instead of using the PIC method, our code employs a continuum kinetic method for its inherent lack of statistical noise and good resolution of distribution functions. These advantages are particularly valuable for studies involving sheaths, as shown by other recent works [46,47,48]. Here, the number of particles per unit phase space volume $f_s(x,v_s,t)$ for each charge species *s* is calculated by advancing the kinetic Boltzmann Transport Equation explicitly over a uniform finite difference 1D-1V grid where the velocity direction is normal to the electrodes ($v_s = v_{s,x}$ in this section).

$$\frac{\partial f_s}{\partial t} = -v_s \frac{\partial f_s}{\partial x} + \frac{\pm q_s}{m_s} \frac{\partial \varphi}{\partial x} \frac{\partial f_s}{\partial v_s} + S_{ch(s)} + S_{coll(s)} \tag{12}$$

The charge source term $S_{ch(s)}$ creates ion-electron pairs with Maxwellian velocity distributions uniformly in space at a rate that adjusts to balance the fluxes of ions $\Gamma_i$ lost to the electrodes. This keeps the mean (spatially averaged) ion density equal to the chosen initial value <N>.

$$S_{ch(s)} = \frac{|\Gamma_{i,x=0}| + |\Gamma_{i,x=L}|}{L} \sqrt{\frac{m_s}{2\pi T_s}} \exp\left(\frac{-m_s v_s^2}{2 T_s}\right) \tag{13}$$

A helium ion mass of 4amu is used throughout this paper. The gap size is L = 30cm throughout. Dirichlet boundary conditions enforce a fixed $V_{bias}$. A flux $\Gamma_{emit}$ of half-Maxwellian electrons is emitted from the cathode. Their temperature $T_{emit}$ is 0.25eV = 2900K throughout.

In discharges, when thermionic or secondary electrons are accelerated by the electric field, they lose energy in inelastic (ionization and excitation) collisions [41]. When the mean free path $e_{mfp}$ is small compared to the system size, a Maxwellian plasma electron population with temperature of a few eV is expected. The general effects of thermalization can be captured using variants of Bhatnagar-Gross-Krook (BGK) [49] collision operators. We use an operator of the form below.

$$S_{coll(e)} = -\frac{|v_e|}{e_{mfp}} f_e + \sqrt{\frac{\pi m_e}{2 T_e}} |v_e| \exp\left(-\frac{m_e v_e^2}{2 T_e}\right) \int_{-\infty}^{\infty} \frac{|v_e| f_e}{e_{mfp}} dv_e \tag{14}$$

Operator (14) ensures that during each time step $\Delta t$ at each $x$, the fraction $|v_e|\Delta t/e_{mfp}$ of electrons with velocity $v_e$ collide and get removed by the negative term. The positive term replaces the collided electrons with an equal number (hence the integral) of thermalized electrons. The thermalized electron distribution is a Maxwellian times $|v_e|$ to compensate for faster particles colliding more often and needing to be replaced at a proportional rate to maintain a plasma with a Maxwellian distribution. Collision operators without the $|v_e|$ dependence can be used to maintain Maxwellian bulk plasma electrons in simulations [25], but the $|v_e|$ proportionality is preferred for this study so that all electrons have a consistent mean free path $e_{mfp}$ independent of their velocity.

For all simulations here, charge source electrons in $S_{ch(e)}$ have $T_e$ = 3eV. The thermalized electron $T_e$ in (14) is also 3eV in the region of space where the mean electron energy exceeds $T_{emit}$, but is set equal to $T_{emit}$ elsewhere. This is to account for the fact that in the region of space before thermoelectrons get accelerated (which is the entire plasma region in the inverse mode) the electrons should have temperature ~$T_{emit}$.

In plasmas, charge-exchange collisions of fast ions (accelerated by the potential) with cold background neutrals act to replace fast ions with cold ions [41]. An ion collision operator analogous to (14) offers a reasonable description of CX collisions. Values of $T_i$ = 0.03eV and $i_{mfp}$ = L/5 = 0.06m are used throughout. Charge source ions $S_{ch(i)}$ also have $T_i$ = 0.03eV.

Our code was run with <N> = 0 and without electron collisions to obtain the vacuum diode solutions in Fig. 1. In the following simulations we will introduce plasma. The parameters {$T_e$, $T_i$, $T_{emit}$, $i_{mfp}$, L} will be held to the aforementioned values. Parameters {$\Gamma_{emit}$, $e_{mfp}$, $V_{bias}$, <N>} will be varied as needed to illustrate essential physics results. In all cases there are 1301 grid points in x, ensuring good spatial resolution of the plasma and sheaths. There are 350 points in velocity space for each species, with separate bounds for ions and electrons to encompass their respective velocity ranges.

## B. Properties of the classical mode

Fig. 5 shows a representative set of classical modes where only $e_{mfp}$ is varied. When $e_{mfp} = L$, some of the thermionic beam transits directly to the anode, as seen in the $f_e$ plot in Fig. 5(c). Cases with $e_{mfp} > L$ are not investigated here because the beam tends to excite strong two-stream instabilities which are studied in Refs. [50,51]. Here in the $e_{mfp} = L$ case, there is a symmetric potential hill structure in the quasineutral region from the presheaths. The presheaths accelerate ions towards the sound speed up to the sheath edges, as evident in the $f_i$ plot.

We observe in Fig. 5(a) that reducing $e_{mfp}$ causes $\varphi(x)$ to become asymmetric in the quasineutral region. This is because a resistivity field is needed to force the current towards the anode against the collisions. It is difficult to isolate the resistivity field since the presheath electric fields are superimposed. The resistivity contribution is inferable from the potential difference between the sheath edges in Fig. 5(a). The potential difference is negligible at large $e_{mfp}$ but consumes a substantial fraction of the bias at low $e_{mfp}$.

Interestingly, the anode sheath potential difference flips sign in the $e_{mfp}/L = 0.03$ case. How is this possible without the anode collecting too many electrons? When the resistivity field is strong, there is a substantial density drop towards the anode in Fig. 5(b) as to be expected by the Boltzmann relation $n_e \propto \exp(-q_e\varphi/T_e)$. The density drop serves to constrain the electron flux reaching the anode.

Next in Fig. 6 we consider a set of cases where only $\Gamma_{emit}$ is varied. The $\Gamma_{emit} = 0$ case has a strong anode sheath since the anode sheath must limit the electron influx to the ion flux lost at the two electrodes. For verification we estimate the total ion flux $2 \times n(T_e/m_i)^{1/2}$ from the Bohm criterion and the anode electron influx $n(T_e/2\pi m_e)^{1/2}\exp(q_e\Phi_{ash}/T_e)$ from a full Maxwellian source. Equating the fluxes predicts an anode sheath potential difference $\Phi_{ash} = -8.5V$, comparable to that observed in Fig. 6(a).

When the cathode injects electrons into the plasma, the anode must collect more electrons for global charge balance. Since $e_{mfp}/L = 0.2$ in Fig. 6, direct transit of thermoelectrons to the anode is inhibited so the anode sheath must weaken to collect more "thermalized" electrons. Because the nonzero values of $\Gamma_{emit}$ in Fig. 6 far exceed the ion flux to the electrodes ($\Gamma_i \sim 10^{17} m^{-2} s^{-1}$) the anode sheath is much weaker than the $\Gamma_{emit} = 0$ case and even flips sign at high enough $\Gamma_{emit}$. We also observe in Fig. 6(a) that raising $\Gamma_{emit}$ leads to the growth of a resistivity field. It is needed to drive a high directed current through the plasma against the collisions which act to isotropize electron velocities.

The $f_e$ slices in Fig. 6(b) confirm that the BGK thermalization operator (14) maintains an electron distribution that is roughly Maxwellian. The suprathermal peaks at x=5cm are from the thermionic beam. Due to sheath acceleration, the beam has a low spatial density relative to the plasma electrons even when $\Gamma_{emit}$ is high.

Another result of raising $\Gamma_{emit}$ is that the electron density increases in the cathode sheath. It is evident in the $\Gamma_{emit} = 3 \times 10^{18}$ m$^{-2}$s$^{-1}$ case in Fig. 6(c) that the total charge in the negative layer is comparable to the positive layer. At higher $\Gamma_{emit}$ values the classical cathode sheath breaks down and current limitation begins. In this example the plasma is able to carry a current about two orders of magnitude larger than the current in the vacuum diode with the same $\{\Gamma_{emit}, T_{emit}, V_{bias}, L\}$ simulated in Fig. 1.

## C. Properties of the conventional SCL mode

Based on conventional hot cathode sheath models without collisions [15], one may expect that raising $\Gamma_{emit}$, reducing $V_{bias}$, or reducing $<N>$ will eventually cause the classical sheath to transition to SCL. Recently it was found that SCL sheaths cannot persist if CX collisions create cold ions that get trapped and accumulate in the VC [25,26]. The ion trapping may be impossible to prevent in experiments, but it is worthwhile to disable it in our simulations so that an equilibrium SCL mode can be studied.

SCL modes in Fig. 7 were obtained by starting from a previous classical mode simulation, turning off ionization and CX collisions just near the cathode, raising $\Gamma_{emit}$ beyond the threshold of VC formation, and letting the plasma settle to equilibrium. One observes that the plasma structure in the SCL modes is similar to a classical mode. In Fig. 7(a) the $f_i$ plot shows that ions accelerate from ~zero velocity near the midplane to supersonic velocities at the cathode. The $f_e$ plot shows that the cathode sheath confines bulk plasma electrons and accelerates the thermionic beam.

Fig. 7(b) shows most of the bias is consumed by the SCL sheath, but there is a potential rise of a few Volts from the cathode sheath edge to anode sheath, indicating a resistivity field. The resistivity field is needed because thermoelectrons that pass the VC accelerate into the plasma and then lose energy by collisions. Unable to overcome the cathode sheath, they must be forced through the plasma into the anode.

We saw in the classical mode (Fig. 6(a)) that raising $\Gamma_{emit}$ necessitated a stronger resistivity field to drive the higher current. In contrast, the SCL $\varphi(x)$ with $\Gamma_{emit} = 10^{20}$ m$^{-2}$s$^{-1}$ in Fig. 7(b) is almost identical to the case with $\Gamma_{emit} = 10^{19}$ m$^{-2}$s$^{-1}$. Once in a SCL mode, further rises of $\Gamma_{emit}$ are offset by the growing VC. The $\Gamma_{emit} = 10^{20}$ m$^{-2}$s$^{-1}$ case in Fig. 7(c) shows a higher $n_e$ close to the cathode but most of those electrons are blocked by the VC and the current is only 6% larger. The saturation is imperfect due to finite $T_{emit}$. The effect of $T_{emit}$ on saturation is illustrated by the vacuum diode simulations in Fig. 1(a).

Incidentally, we note that the anode sheath potential difference is negative but very weak in Fig. 7(b). We verified that it can flip to positive if $e_{mfp}$ is reduced, as in the classical mode. We conclude that the anode sheath potential difference can take either sign when the cathode sheath is classical or SCL.

**D. Properties of the Inverse Mode**

If the ion CX collisions are included near the cathode, then as soon as a SCL sheath starts to form, a cloud of trapped CX ions will accumulate in the VC and spread across the plasma, forcing a transition to the inverse mode. This temporal transition was studied in Ref. [26]. The ion mean free path $i_{mfp}$ affects the rate of the transition but the transition will occur for any finite $i_{mfp}$. As the inverse mode forms, CX collisions deplete whatever extra kinetic energy trapped ions acquired during the transition. When an equilibrium inverse mode is reached, a cold background of ions remains and the particular $i_{mfp}$ is unimportant.

Equilibrium inverse modes are shown here in Fig. 8 to compare to our theoretical picture. As expected, $\varphi(x)$ in the plasma is flat and below both electrode potentials. The full bias is consumed by the anode sheath. The $f_e$ plots show that electrons accelerate through the anode sheath. The $f_i$ plots show that ions are trapped between the sheaths. Ion losses to the cathode are not exactly zero but are low enough that the offsetting volumetric charge source does not affect the mode structure.

The only parameter varied among the Fig. 8 cases is $e_{mfp}$. Even as the collisionality is raised by 500 times, no noticeable potential difference forms between the sheath edges. This is a stark contrast from the separation of the sheath edge potentials seen in the classical mode in Fig. 5(a). The distinct behavior is traced back to the fact that thermoelectrons entering the plasma can easily return to the cathode in the inverse mode, so no potential rise from "resistivity" is needed. All of the trapped cold ions will settle to the lowest potential, so the entire plasma must be an equipotential in equilibrium, regardless of $e_{mfp}$.

The profound effect of collisionality on the plasma density distribution is evident in Fig. 8. In the case with $e_{mfp}/L = 3.33$, the density is almost uniform in space. Most thermoelectrons in the plasma transit collisionlessly to the anode (most have $v_e > 0$ in the $f_e$ plot). In the more collisional cases the plasma density drops more towards the anode sheath edge. In all cases the density profile appears roughly linear.

| $e_{mfp}/L$ | $\Gamma_e/\Gamma_{r,(x=1cm)}$ (theory) | $\Gamma_e/\Gamma_{r,(x=1cm)}$ (simulation) | $N_{cse}/N_{ase}$ (theory) | $N_{cse}/N_{ase}$ (simulation) |
|---|---|---|---|---|
| 3.33 | 0.87 | 0.87 | 1.30 | 1.38 |
| 0.33 | 0.40 | 0.41 | 4.00 | 3.42 |
| 0.033 | 0.063 | 0.079 | 31.0 | 6.64 |
| 0.0067 | 0.013 | 0.017 | 151 | 15.5 |

TABLE I. The effects of collisions on the inverse mode in simulations is compared to theory. $\Gamma_e/\Gamma_{r,(x=1cm)}$ represents the flux of electrons reaching the anode compared to the rightward flux injected into the neutral region. We note that when electrons move only in the x-direction, the theoretical formulas (10) and (11) should contain an extra factor of two in front of $e_{mfp}$. So the current attenuation factor from collisions in simulations is predicted to be $\approx 1/(1+L/(2e_{mfp}))$, approximating $L_p = L$. The predicted ratio of densities at the sheath edges is $N_{cse}/N_{ase} \approx (L/e_{mfp} + 1)$.

    Our Sec. III.B gave a first-principles derivation for the linear plasma density distribution and the current attenuation by collisions. The accuracy of formulas (10) and (11) is tested in Table I. The formulas show decent agreement with simulation results at modest collisionality. The predicted ratio $N_{cse}/N_{ase}$ becomes inaccurate when $e_{mfp}$ is very small. The discrepancy might be from the fact that the anode sheath becomes highly collisional and the theoretical treatment of its sheath edge as a perfectly absorbing boundary to plasma electrons breaks down.

    The influence of other simulation parameters on the inverse mode is worth discussing. Starting from the equilibrium case with $e_{mfp}/L = 0.0067$ from Fig. 8, $V_{bias}$ was slowly adjusted to a new value and the plasma was allowed to reach a new equilibrium. The results in Fig. 9 indicate that $V_{bias}$ only affects the potential of the anode sheath. This is expected since the potential everywhere else is constrained by the plasma density and emission properties through Eq. (9).

    Fig. 10 shows how the plasma and sheaths in the inverse mode change when the mean ion density $<N>$ is varied over four orders of magnitude. In the lowest density case, the neutral region is thin. The $\varphi(x)$ looks like the current-limited Child-Langmuir $\varphi(x)$ with a slightly extended potential minimum. Conventional wisdom suggests that the presence of ions should raise the current flow beyond the CL threshold. However, for three of the cases in Fig. 10 the net electron flux $\Gamma_e$ is actually lower than the net flux in the vacuum diode in Fig. 1 with the same $\Gamma_{emit}$ ($10^{19} m^{-2} s^{-1}$), $T_{emit}$ (0.25eV), $V_{bias}$ (15V), and L (30cm). This demonstrates how dramatic the current attenuation by collisions can be in the inverse mode.

    The inverse sheath is weaker in the higher plasma density cases in Fig. 10. If too much plasma is added, the inverse sheath collapses and the mode breaks down. The density limit is captured in Eq. (9) but its physical origin is more apparent in Fig. 10. In the highest $<N>$ case, the plasma density at the cathode sheath edge is close to $n_e$ at the cathode. It is clear that if the density of ions were raised too much, the hot cathode could no longer provide enough thermoelectrons to neutralize them.

    We emphasized that the electric field in an inverse mode plasma is zero in the "cold ion limit". But what if $T_i > 0$? The plasma density gradient combined with a nonzero ion temperature will form an ion pressure gradient $d(n_i T_i)/dx$. That will require an electric field to maintain equilibrium on the ions, which are confined and must have zero mean (flow) velocity. This idea is tested by simulations in Fig. 11. The case with $T_i$ of 0.03eV, typical when ions are in equilibrium with a room temperature neutral gas, is close enough to the cold ion limit that the ion pressure and electric field are negligible. In practice, higher $T_i$ values are possible if heavy particles are heated perhaps by the hot cathode or by electron impacts. The higher $T_i$ cases in Fig. 11(a) exhibit a noticeable electric field in the direction expected to offset the ion pressure gradient. The field happens to have the same sign as the "resistivity" fields demonstrated earlier for the classical modes (Figs. 5,6). But the field's purpose here is profoundly different and its strength is sensitive to different parameters. We also note that the electric field makes the plasma density distribution

deviate to a slight extent from its ideal linear form, see Fig. 11(b). But overall, the ion temperature effects do not change the nature of the inverse mode.

## V. INSTABILITIES IN THE INVERSE MODE

Plasma diodes are subject to a variety of instability phenomena [26,27,28,51,52,53,54] which can affect the plasma properties and device performance. Our simulations find that plasmas in the inverse mode are sometimes stable and sometimes unstable. Of the cases in Fig. 8, only the $e_{mfp}/L = 0.0067$ case is truly static. The data for the unstable cases represent the plasmas when they are closest to equilibrium, before an instability begins. At that time, the plasmas are well-described by our theoretical model in Sec. III.

### A. Instability Dynamics

Fig. 12 shows the time evolution of the $e_{mfp}/L = 0.33$ case. Instability begins when a small double layer (DL) is born within the plasma in Shot 1. Instability dynamics differ from run to run depending on simulation parameters. Generally, the DL's amplify in time while propagating towards the anode. Sometimes multiple DL's are present at once, as in Shot 3. The DL's accelerate ions towards the cathode, some of which may escape. CX collisions re-trap ions and cool them again. By Shot 6, it is evident that the original equilibrium has been restored. Then another instability occurs.

The instabilities seen here are related to the "self-oscillations" observed in simulations and experiments of AGM thermionic discharges [28,44,54] and single-ended Q machines [55,56]. The cathode sheath is inverse in those situations. In the previous studies, the self-oscillations were shown to feature moving DL's that push ions closer to the cathode, leading to a temporary increase of ion density there and a temporary increase of circuit current.

It was not previously explained how much current flows in the inverse mode and how the plasma distributes itself. Our model gives the first theoretical foundation for the inverse mode properties in its equilibrium, before DL's form. States with moving DL's can be thought of as distortions to the equilibrium, but their dynamics are complicated. As shown in Fig. 13, the shape of the current oscillations varies substantially among cases. As the oscillation time scale depends on the movement of slow trapped ions, it is difficult to run simulations long enough to know if they ever reach a true periodic oscillation. Interestingly the $e_{mfp}/L = 0.033$ case appears to be in a regular oscillation for many periods but it switches to another oscillation regime at around $t = 150\mu s$.

### B. Conditions for stability and instability

Although the DL instability was well studied in the past, its cause is not fully understood. We discovered that its formation depends on electron collisionality. As shown in Fig. 13, the current is stable (temporally constant) in the lowest $e_{mfp}/L$ case. Increasingly intense oscillations of relative maximum-to-minimum current are seen at higher $e_{mfp}/L$ values. From our inspection, the relationship between collisions and instability appears to be from the fact that the requisite space charge layers of the DL can form more easily if the initial background distribution of electrons is sufficiently "lopsided", meaning more electrons move right than left. The distribution is more lopsided when collisions are rare, as expected from theory and shown in Fig. 14.

Growing DL instabilities of this type are not seen in classical and SCL modes. One possible reason is that the interior $f_e$ is never lopsided; the confined plasma electrons are roughly Maxwellian and the tail produced by the thermionic beam makes just a small distortion, see Fig. 6(b). Also, it appears necessary for the ions to be cold and non-flowing in order for the DL instability to excite.

## VI. FURTHER DISCUSSION OF THE INVERSE MODE

### A. Example Applications

Today, SCL sheaths are still assumed to form under strong thermionic emission in numerous contexts. Examples include tokamak divertor plates [57], hollow cathode neutralizers [58,59], emissive probes [9], spacecraft tethers [8], and transpiration cooling of hypersonic vehicles [12]. The existence of inverse modes may enable operating regimes with practical advantages or drawbacks not previously recognized. The set of applications is too diverse to discuss each one in detail. Some general implications will be mentioned here.

### B. Cathode Sputtering Differences Among Current Modes

Hot cathode sputtering by ion impacts limits the lifetime of many devices [5,58]. Sputtering is often thought to be an inescapable problem, since acceleration of ions into the cathode is inevitable under the prevailing assumption that the cathode sheath must be classical or SCL.

Designing devices to operate in the inverse mode could minimize or eliminate sputtering. Not only is the flux of plasma ions that overcome the inverse sheath low or negligible, their energies will be too low to cause sputtering. If ions are created in the anode sheath, they could get accelerated towards the cathode. But to cause sputtering they would have to transit the gap and their energy may dissipate in collisions. Also, in devices it may be possible to direct the anode surface normal away from the cathode.

### C. Current Flow and Input Power Optimization Among Current Modes

Under the conventional assumption that a hot cathode sheath must be classical or SCL, a strong bias is often needed to drive the desired current, thereby raising the required power. In the inverse mode the current flow is independent of $V_{bias}$, for a given plasma density. A given current could be transmitted at very low power in principle.

A drawback of the inverse mode is that the current will be less than the available emission. Although we noted the entire thermionic flux could flow through a plasma in the inverse mode if it is collisionless and its density approaches $N_{emit}$, the instability discussed in Sec. V disrupts such a state. We found that inverse modes could be stabilized by introducing sufficient neutral gas to make $e_{mfp}/L_p \ll 1$, but the same collisions will attenuate the current according to Eq. (10). So it would be worthwhile to investigate whether other factors such as multidimensional shear effects may suppress the instability in the weakly collisional regime. Simultaneous high current, low power operation would then be possible.

### D. Spatial Power Dissipation in Inverse Versus Other Modes

Quantitative calculation of power dissipation in a plasma facing a hot cathode is a challenging task [60] requiring consideration of collisions among electrons, ions and neutrals. Some general comparisons can be made among the current modes based on their potential distributions. In classical and SCL modes, much of the power is deposited when the accelerated thermoelectrons suffer inelastic and ionization collisions in the plasma volume. Plasma electrons undergo Joule heating by the resistivity field. Both electrodes are heated by particle bombardment. Ion bombardment heating of the cathode is often enough to self-sustain the thermionic emission in arcs [41].

In the inverse mode, since thermoelectrons do not gain any energy until they reach the anode sheath, most of the power should be deposited there or on the anode. This could benefit processing applications where anode vaporization is desired and cathode vaporization is not. Ideally, the cathode in the inverse mode suffers no heating from ions and undergoes evaporative cooling due to the loss of thermoelectrons. Hence external cathode heating is needed to sustain an inverse mode. Since ions are confined, little power is needed to sustain the plasma. Ion lifetime is set by slow loss processes such as volumetric recombination, multidimensional diffusion, or transient ejection by DL instabilities if present.

### E. Improving Future Simulation Models of Emitting Sheaths

Our simulations show that it is crucial to include ion collisions in any future simulation studies that involve strongly emitting surfaces. The SCL and inverse modes in Figs. 7 and 8 have similar control parameters, yet they are very different states of plasma. The only difference between the setups is that the ionization and CX collisions were absent from a thin region near the cathode in the SCL to prevent destruction of the mode by ion trapping. Inverse sheaths are expected to prevail in experiments since some neutrals will always be present in plasma sheaths to enable CX collisions.

The idea that ion trapping inhibits SCL sheaths explains why SCL sheaths were only seen in simulations that omitted collisions such as Refs. [12,15,20,57]. SCL sheaths do not appear in simulations that include collisions. For example, in the seminal simulation studies of thermionic discharges [44] the current-limited mode was the AGM with an inverse sheath ("Langmuir modes" with SCL sheaths were not seen). Another example is in the PIC simulations of Hall discharges of Ref. [61], where the sheaths switched from classical to inverse (never SCL) when the emission coefficient at the floating channel walls was raised from below unity to above.

### F. Inverse Modes in Contact Ionization Sources

Our simulation study considered plasma diodes where the ions are produced volumetrically within the electrode gap. In some applications such a Q machines [55] and thermionic converters [62], the ions are created by contact ionization at the cathode. It is worthwhile to mention that the model of inverse mode presented here could carry over to describe the plasma structure in contact ionization sources since an inverse sheath naturally allows ions born at the cathode into the plasma. Interestingly, a current-limited mode in recent thermionic converter simulations appears to be inverse, with a flat plasma potential below the cathode potential, see the "0%" case in Fig. 5 of Ref. [62]. It would be inconceivable for a SCL sheath to form in a current-limited state of a contact ionization source since the sheath would block the ion injection, preventing the creation of the interior plasma.

## G. Experimental identification of current-limited modes

In experimental work, it was often assumed that a hot cathode sheath is SCL when the measured current is less than the available thermionic emission. Since inverse sheaths can also limit current, additional information is needed to identify which current-limited mode is present. Although inverse modes are now expected to prevail over SCL due to the ion trapping effect, it is worthwhile to discuss how to conclusively distinguish these modes by measurement.

A discussion of past experiments involving floating surfaces under strong electron emission in Sec. V of Ref. [63] showed that most of the early experimental tests of SCL sheath theory were either inconclusive or revealed discrepancies that are more consistent with inverse sheath theory. To our knowledge there are no measurements proving the presence of a SCL sheath at a biased strongly emitting surface either. Researchers in Ref. [64] offered emissive probe measurements of a nonmonotonic $\varphi(x)$ near a negatively biased electrode and suggested it is a virtual cathode caused by secondary emission by ion impacts. However, the accuracy of probe measurements in the sheaths of other objects is suspect and the present author doubts that ion-induced emission can be intense enough to form a VC.

Other techniques might be used to identify the current mode without sheath measurement. Emissive probe measurements in the plasma interior were used to identify AGM's in thermionic discharges [27,28]; the fact that the measured plasma potential was close to and below the cathode potential proves that the cathode sheath was inverse (if it were SCL the plasma potential would have to be far higher, near the anode potential). Light emission profiles were also used to deduce that the plasma potential was low [27], but light will not be produced in applications where the bias is too weak for thermoelectrons to reach excitation energies. Laser-induced fluorescence [65] measurements of ion velocities might then be used to determine whether ions are accelerating into the cathode or not. Also, if measurable oscillations of current or other parameters are present, their frequency may help identify the mode. The beam-plasma instabilities in the inverse mode have frequencies orders of magnitude lower than the Langmuir frequency two-stream instabilities [51] that may arise if the cathode sheath is classical or SCL.

## H. Extending the planar inverse mode to other geometries

SCL sheath solutions in curved geometry have been treated [66] to model thin filament cathodes. Inverse modes are also possible in curved geometry situations. Convincing evidence is in the seminal thermionic discharge study by Malter et al. [27] which used a cylindrical concentric electrode configuration. Their probe measurements in the AGM confirmed that the plasma electron temperature was close to the inner hot cathode temperature, and the potential was below the cathode potential over most of the gap except in the outer anode sheath. These properties are like a planar inverse mode. A significant difference anticipated for cylindrical geometry is in the current flow; the thermoelectrons undergoing the random collisional walk have a higher probability of reaching the anode compared to the planar case.

The fundamental premises of our inverse mode theory can be extended to arbitrary geometric configurations. Thermoelectrons injected from the cathode sheath edge enter a zero-field region influenced only by collisions, and get lost when they reach either boundary (cathode sheath edge or anode sheath edge). Monte Carlo algorithms tracking the trajectories of an ensemble of random walking particles could be used to calculate the electron density distribution in space and the fraction of electrons reaching the anode. Assuming ions are trapped and damped by collisions with cold neutrals, they will prohibit any interior potential gradients and settle to the same density distribution as the electrons. However, the presence of plasma sources or losses to boundaries other than the electrodes could introduce nontrivial flows or potential gradients that complicate the dynamics.

## VII. Conclusions

We presented an alternative theoretical framework describing how a hot cathode, plasma and anode can interact when the thermionic current is limited. In the "inverse" mode the cathode sheath, anode sheath, presheaths, resistivity field, plasma temperature, electrode erosion rate, and dissipated power distribution will drastically differ from conventional models of space-charge limited flow. Analytical expressions for the plasma properties, sheath properties and net current in a planar geometry inverse mode were derived.

A new simulation code was developed to study the thermoelectron flow through a plasma. Our code used a noise-free continuum kinetic method, simulated the entire electrode gap, and allowed intrinsic plasma properties to be varied independently. This enabled the physical factors governing the potential distribution between the electrodes to be studied in unprecedented detail. The coupled influence of the cathode sheath, presheaths, resistivity field, and anode sheath was clarified for the classical, conventional SCL, and inverse current modes. Data for the inverse mode was in good quantitative agreement with our theoretical model. Our 1D planar theoretical and simulation models could be extended to multidimensional configurations in the future.

The existence of inverse modes is experimentally validated by past measurements of thermionic discharges in the anode glow mode. But inverse modes are possible in a wide variety of other devices with hot cathodes. Knowing that charge-exchange ion trapping inhibits SCL sheaths, the inverse mode should be the only current-limited equilibrium realizable. In the future, some hot cathode devices might be designed to exploit certain unique properties of inverse mode.

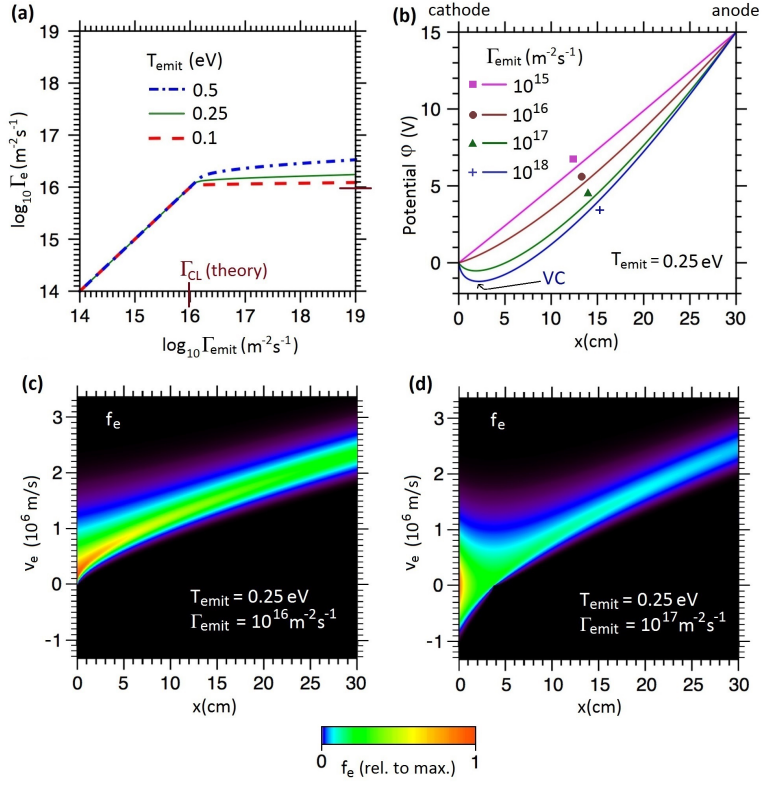

FIG. 1. Demonstration of space-charge limitation in a planar vacuum diode with gap length 30cm and $V_{bias}$ = 15V. Panel (a) shows the net electron flux $\Gamma_e$ as $\Gamma_{emit}$ is varied over five orders of magnitude, for three separate $T_{emit}$ values. The maximum $\Gamma_e$ predicted by the CL formula [1] with $T_{emit} = 0$ is $\Gamma_{CL} = 9.4\times10^{15}$m$^{-2}$s$^{-1}$. Panel (b) shows the evolution of $\varphi(x)$ and formation of a VC as $\Gamma_{emit}$ is raised. Distribution functions $f_e$ of the density of electrons in $(x,v_e)$ space are shown for a representative temperature-limited state (c) and space-charge limited state (d). To model a plasma diode, Langmuir extended the CL solution by introducing positive ions flowing opposite to the thermoelectrons. Our paper presents an alternative solution where trapped ions are placed at the potential minimum of a VC.

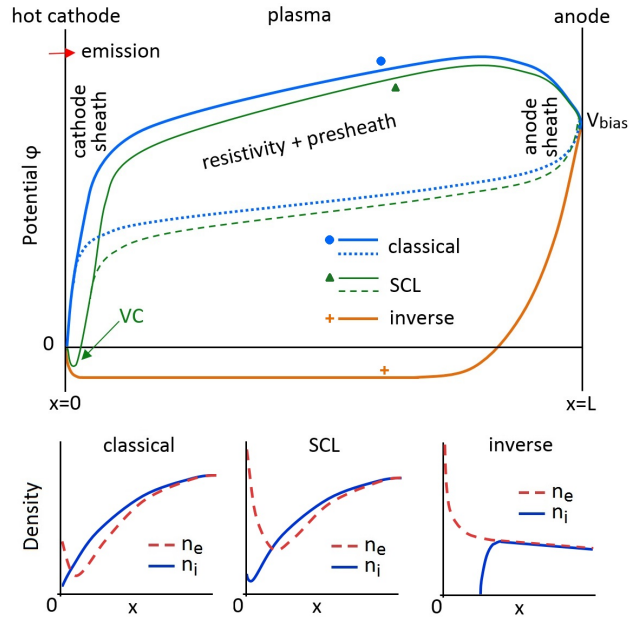

FIG. 2. Schematic of the possible potential distributions in a plasma diode with volumetric sourcing of plasma particles in the gap. The structure of the charge layers in each type of cathode sheath is sketched. In modes with classical or SCL cathode sheaths, a resistivity field is expected and the anode sheath potential difference can take either sign. In the inverse mode, there can be no resistivity field.

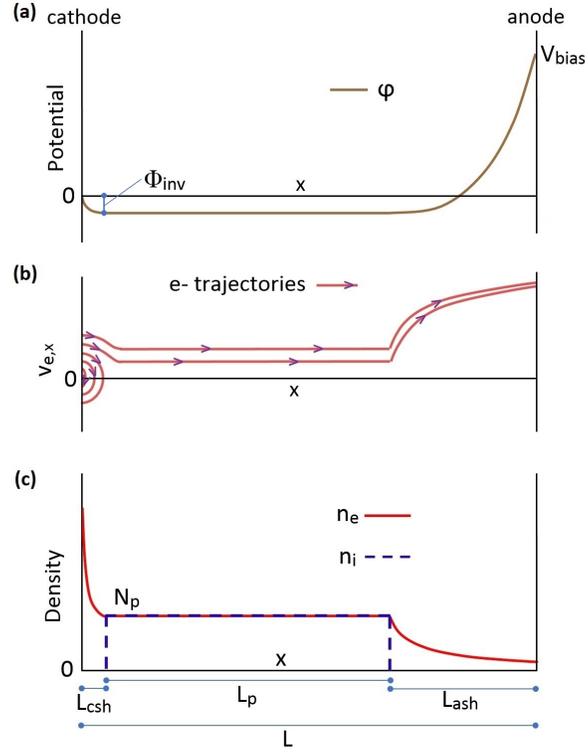

FIG. 3. Qualitative sketch of the (a) potential distribution, (b) electron trajectories in phase space, and (c) charge density distributions in the collisionless inverse mode.

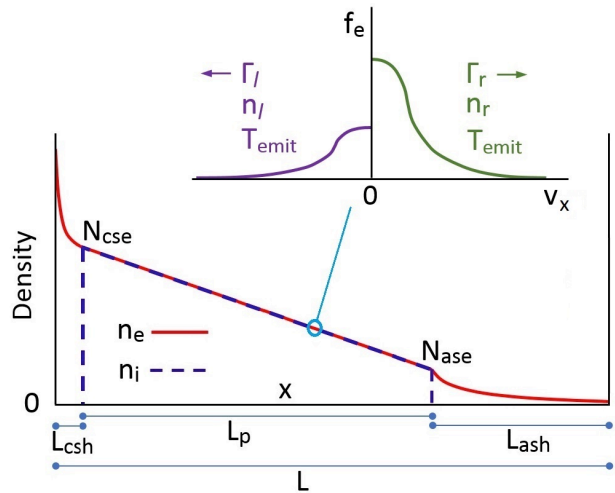

FIG. 4. Sketch of the charge density distribution in the collisional inverse mode. At each x in the plasma, the electron velocity distribution is two half-Maxwellians of temperature $T_{emit}$ with densities $n_r$ and $n_l$. Our model calculates the rightward and leftward electron components consistently accounting for their redirection by collisions and losses through the sheaths. The potential distribution is similar to $\varphi(x)$ in the collisionless case (Fig. 3(a)).

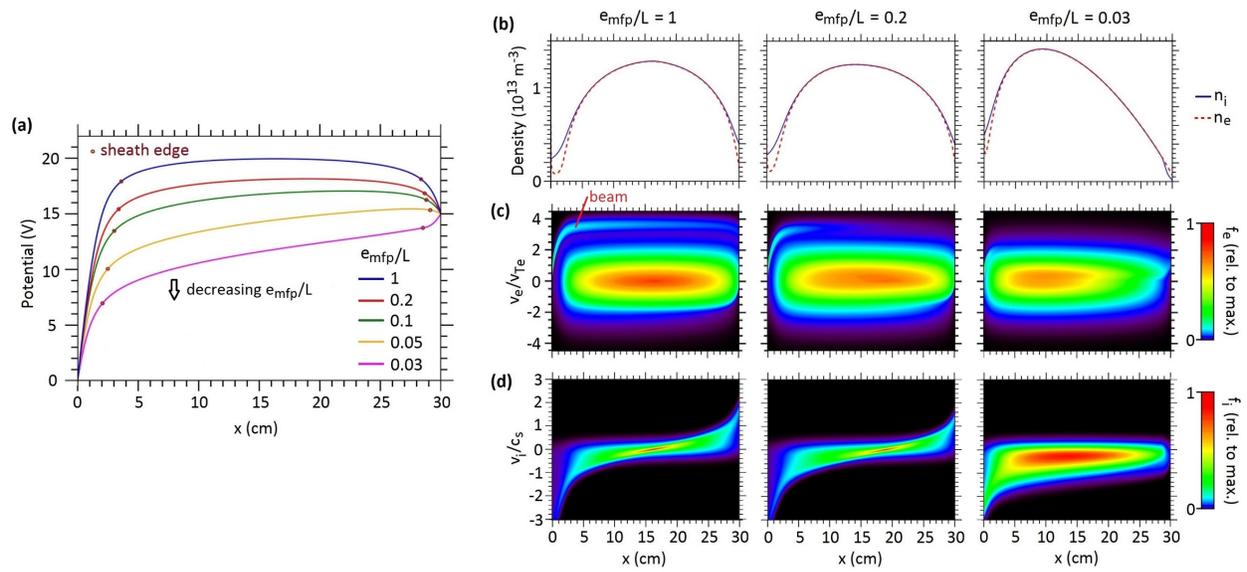

FIG. 5. Effect of electron collisions on the plasma and sheaths in the classical current mode. Fixed parameters include $\{<N> = 10^{13} m^{-3}, \Gamma_{emit} = 10^{18} m^{-2} s^{-1}, V_{bias} = 15V\}$. Panel (a) shows $\varphi(x)$ in five cases with the electron thermalization mean free path $e_{mfp}$ ranging from 3% to 100% of the electrode gap. For three of these cases we show the (b) charge density profiles, (c) electron distribution functions (note the presence of an accelerated thermionic "beam" separate from the bulk plasma electrons), *and* (d) ion distribution functions. Electron velocities are normalized to the thermal velocity $v_{Te} \equiv (T_e/m_e)^{1/2} = 726$ km/s. Ion velocities are normalized to the sound speed $c_s \equiv (T_e/m_i)^{1/2} = 8.48$ km/s. The sheath edges in our figures are defined as the point where $n_e$ differs from $n_i$ by 5%.

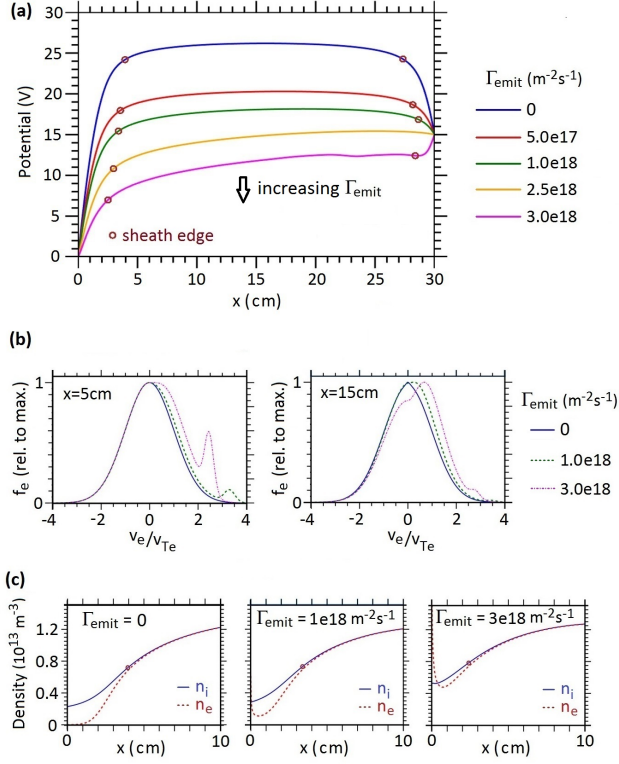

FIG. 6. Effect of the emitted flux on the classical current mode. Fixed parameters include $\{<N> = 10^{13} m^{-3}, V_{bias} = 15V, e_{mfp}/L = 0.2\}$. Panel (a) shows the potential distributions in five cases with different $\Gamma_{emit}$. Panel (b) shows slices of the electron distribution functions at x=5cm and at the midplane. Panel (c) shows the charge density distributions in the cathode sheath, demonstrating the formation of a negative charge layer as $\Gamma_{emit}$ is raised. We note that the $\Gamma_{emit} = 3\times10^{18}$ m$^{-2}$s$^{-1}$ case is unstable and undergoes a relaxation oscillation that is not yet understood. Cases with positive anode sheath potential difference sometimes feature this instability but not always (the $e_{mfp}/L = 0.03$ case in Fig. 5 is stable).

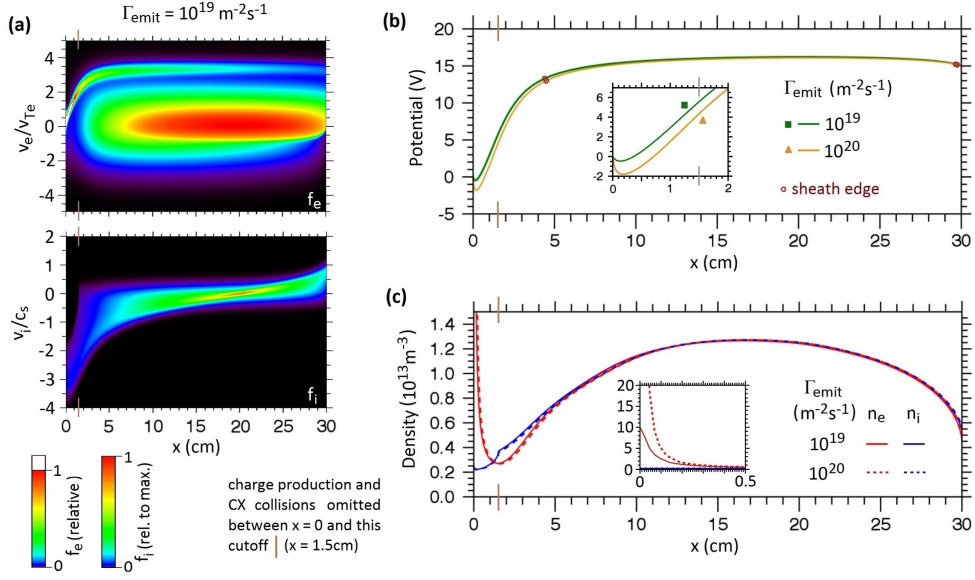

FIG. 7. Demonstration of the space-charge limited current mode with a virtual cathode as predicted by conventional theories. Parameters include {$e_{mfp}/L = 0.5$, $V_{bias} = 15V$, $\langle N \rangle = 10^{13}$ m$^{-3}$}. Panel (a) shows the distribution functions when $\Gamma_{emit} = 10^{19}$ m$^{-2}$s$^{-1}$. The (b) potential and (c) charge density distributions are shown for the $\Gamma_{emit} = 10^{19}$ m$^{-2}$s$^{-1}$ simulation and a similar case with $\Gamma_{emit} = 10^{20}$ m$^{-2}$s$^{-1}$. The net flux $\Gamma_e$ in these cases is $3.02\times10^{18}$ m$^{-2}$s$^{-1}$ and $3.21\times10^{18}$ m$^{-2}$s$^{-1}$, respectively. The cutoff is introduced to prevent the accumulation of trapped ions in the VC, which destroys SCL sheaths and forces a transition to the inverse regime [26].

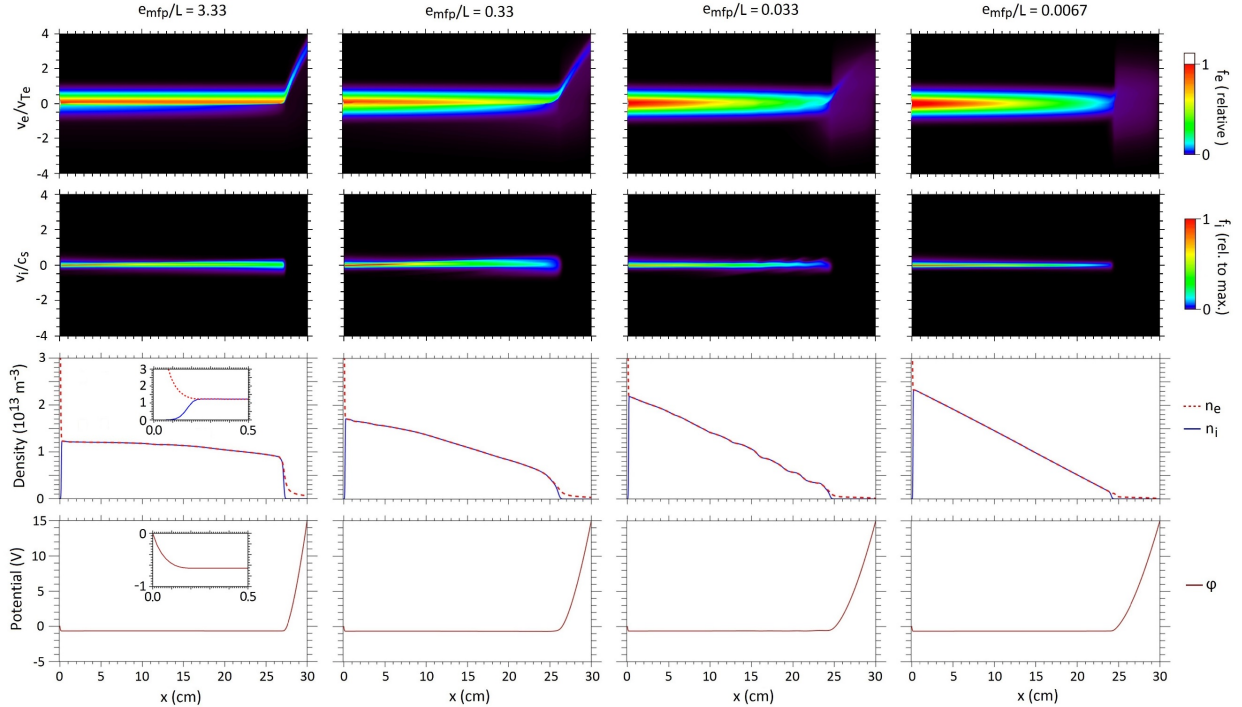

FIG. 8. Representative simulations of the inverse mode with four different $e_{mfp}$ values. Fixed parameters including {$\langle N \rangle = 10^{13}$ m$^{-3}$, $\Gamma_{emit} = 10^{19}$ m$^{-2}$s$^{-1}$, $V_{bias} = 15V$} are identical to the SCL mode of Fig. 7(a). The only difference in the setup here is that the ionization and CX collisions are included near the cathode.

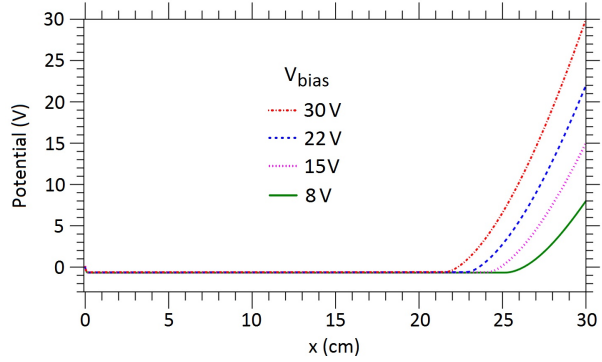

FIG. 9. Comparison of inverse modes with different $V_{bias}$. All simulation parameters besides $V_{bias}$ are equal to the case in Fig. 8 with $e_{mfp}/L = 0.0067$

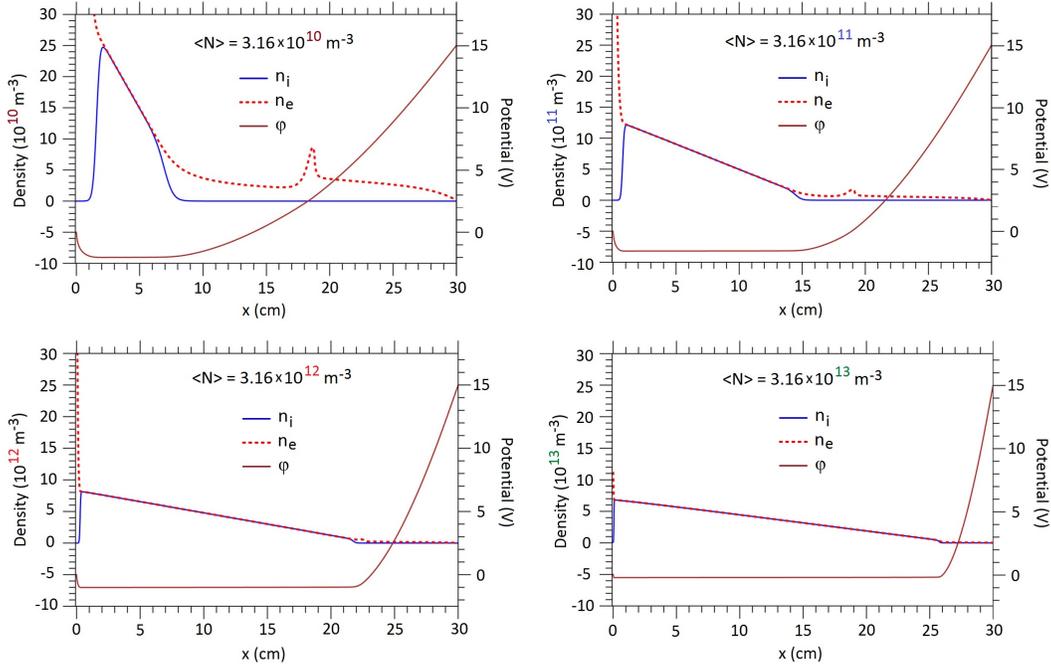

FIG. 10. Comparison of inverse modes with different plasma densities. All parameters besides $<N>$ are equal to the case in Fig. 8 with $e_{mfp}/L = 0.0067$. The net electron fluxes in the four cases here are $\{1.28\times10^{15} m^{-2}s^{-1}, 2.88\times10^{15} m^{-2}s^{-1}, 1.30\times10^{16} m^{-2}s^{-1}, 9.25\times10^{16} m^{-2}s^{-1}\}$, in order of increasing $<N>$. We note the peaks of $n_e$ in the anode sheath are attributable to the thermalization operator (14). It was configured to thermalize electrons to $T_{emit}$ when the mean electron energy is below $T_{emit}$, and otherwise to $T_e = 3eV$. So a distortion to $n_e$ occurs where the electrons surpass the threshold energy in the anode sheath.

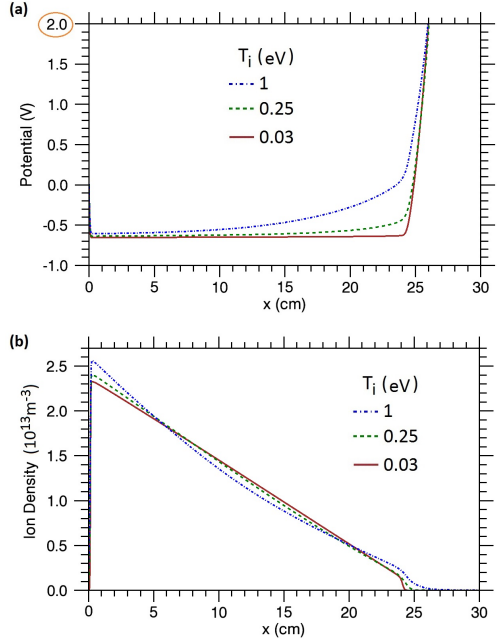

Fig. 11. Effect of ion temperature on the potential (a) and density (b) distributions in the inverse mode. Other simulation parameters are equal to the Fig. 8 simulation with $e_{mfp}/L = 0.0067$. In the $T_i = 0.03$eV case, the distributions take the linear forms expected from the theoretical model in the cold ion limit. Higher ion temperatures distort the distributions to a minor extent.

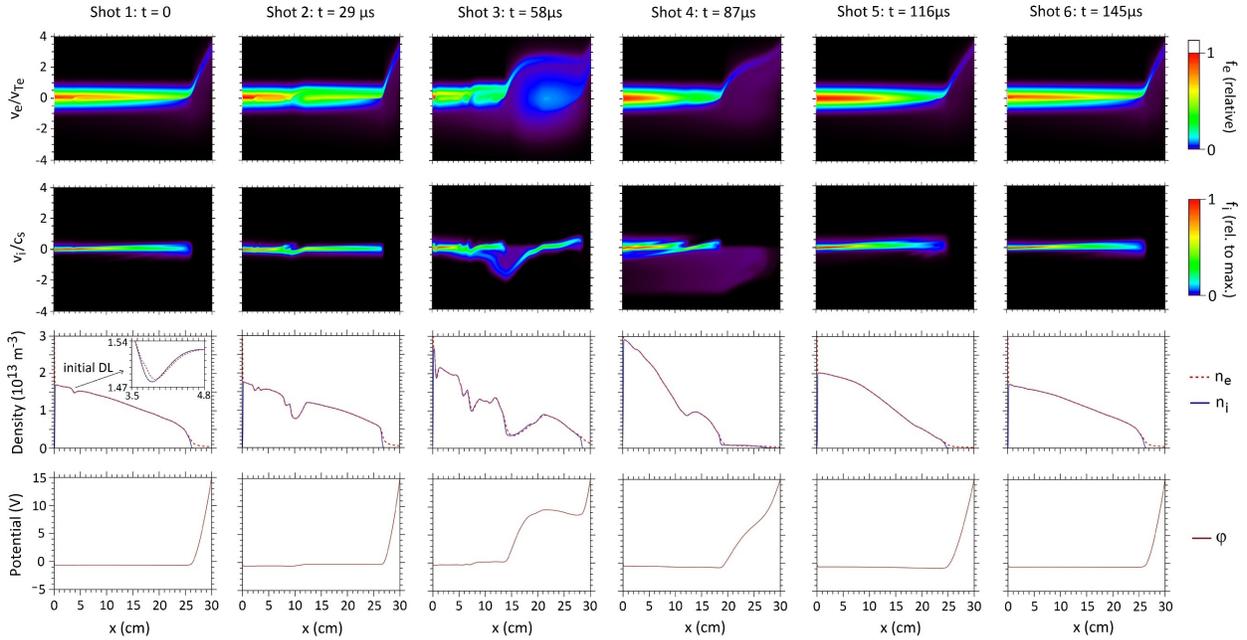

FIG. 12. Dynamics of an inverse mode instability from the Fig. 8 simulation with $e_{mfp}/L = 0.33$. Snapshots show the plasma every 29μs from the initial onset of instability until the plasma returns to a near-equilibrium state.

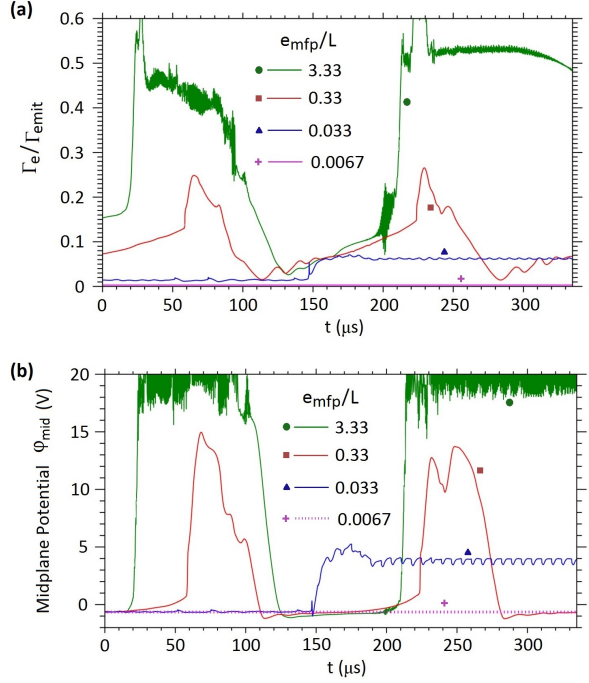

FIG. 13. Time evolution of the (a) net electron flux normalized to $\Gamma_{emit}$ *and* (b) the midplane potential relative to the cathode. The t=0 states for each $e_{mfp}$ case are the states shown in Fig. 8. While $\varphi_{mid} < 0$, the system is close to an inverse mode equilibrium. When $\varphi_{mid} > 0$ a DL is present and temporary increases of current are observed. The high frequency oscillations seen in some intervals of the $e_{mfp}/L = 3.33$ case are two-stream instabilities enabled when the DL accelerates the thermionic beam to high velocity.

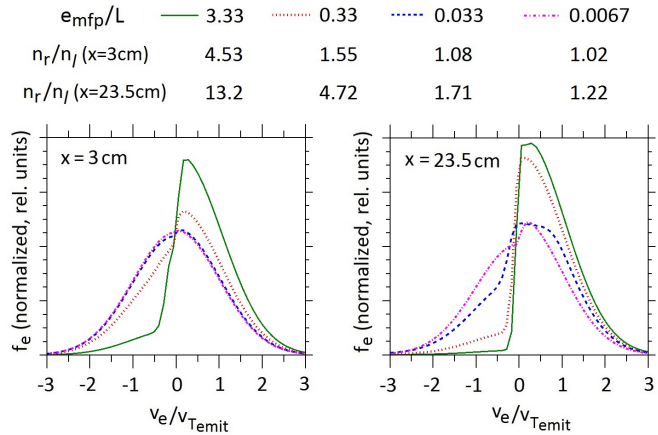

FIG. 14. Slices of the electron velocity distribution at x = 3cm and x=23.5cm in the initial "equilibrium" states (without DL's). The curves are normalized to total electron density so that each curve has the same area. This figure intends to compare how lopsided $f_e$ is near the boundaries of the neutral region for each $e_{mfp}$ case. The ratio of the density of electrons moving right and left is a metric of lopsidedness. Near the cathode sheath, theory suggests $n_r/n_l \to \infty$ when $e_{mfp}/L \to \infty$ and $n_r/n_l \to 1$ when $e_{mfp}/L \to 0$. DL instabilities are more prevalent at larger $n_r/n_l$. The velocity axis is normalized to the thermoelectron thermal speed $v_{Temit} \equiv (T_{emit}/m_e)^{1/2} = 210$ km/s.

**Acknowledgment:**

This work was performed under the auspices of the U.S. Department of Energy by Lawrence Livermore National Laboratory under Contract No. DE-AC52-07NA27344, and supported by U.S. DoE, Office of Science, Fusion Energy Sciences. Funding was also provided by the DOE Basic Plasma Science project "Effects of Electron-Emitting Surfaces on Plasma Properties".